\documentclass[aps,prd,preprint,floatfix,superscriptaddress,nofootinbib,longbibliography,a4paper]{revtex4-1}
\pdfoutput=1
\usepackage{color}
\usepackage{graphicx}
\usepackage{textcomp}
\usepackage{subfigure}
\usepackage{feynmp}
\usepackage{amsmath,amssymb,array}
\usepackage{enumerate}
\usepackage{slashed}
\usepackage{hhline}
\usepackage{hyperref}
\hypersetup{colorlinks=true,
	linkcolor=blue,
	filecolor=magenta,      
	urlcolor=blue}
\usepackage{tcolorbox}
\definecolor{bred}{rgb}{1.0,0.0,0.0}
\definecolor{bgreen}{rgb}{0.0,1.0,0.0}
\newtcbox{\rb}{on line,
  colframe=bred,colback=bred!15!white,
  boxrule=0.0pt,arc=3pt,boxsep=0pt,left=2pt,right=2pt,top=2pt,bottom=2pt}
\newtcbox{\gb}{on line,
  colframe=bgreen,colback=bgreen!15!white,
  boxrule=0.0pt,arc=3pt,boxsep=0pt,left=2pt,right=2pt,top=2pt,bottom=2pt}

\newcommand{\beqa}{\begin{eqnarray}}
\newcommand{\eeqa}{\end{eqnarray}}
\newcommand{\be}{\begin{equation}}
\newcommand{\ee}{\end{equation}}
\newcommand{\ba}{\begin{array}} 
\newcommand{\ea}{\end{array}}

\begin{document} 
\vspace*{0.5cm}
\title{Gauged $SU(3)_F$ and loop induced quark and lepton masses }
\bigskip
\author{Gurucharan Mohanta}
\email{gurucharan@prl.res.in}
\affiliation{Theoretical Physics Division, Physical Research Laboratory, Navarangpura, Ahmedabad-380009, India}
\affiliation{Indian Institute of Technology Gandhinagar, Palaj-382055, India\vspace*{0.5cm}}
\author{Ketan M. Patel}
\email{kmpatel@prl.res.in}
\affiliation{Theoretical Physics Division, Physical Research Laboratory, Navarangpura, Ahmedabad-380009, India}

\begin{abstract}
We investigate a local $SU(3)_F$ flavour symmetry for its viability in generating the masses for the quarks and charged leptons of the first two families through radiative corrections. Only the third-generation fermions get tree-level masses due to specific choice of the field content and their gauge charges. Unprotected by symmetry, the remaining fermions acquire non-vanishing masses  through the quantum corrections  induced by the gauge bosons of broken $SU(3)_F$. We show that inter-generational hierarchy between the masses of the first two families arises if the flavour symmetry is broken with an intermediate $SU(2)$ leading to a specific ordering in the masses of the  gauge bosons. Based on this scheme, we construct an explicit and predictive model and show its viability in reproducing the realistic charged fermion masses and quark mixing parameters in terms of not-so-hierarchical fundamental couplings. The model leads to the strange quark mass, $m_s \approx 16$  MeV at $M_Z$, which is $\sim 2.4 \sigma$ away from its current central value. Large flavour violations are a generic prediction of the scheme which pushes the masses of the  new gauge  bosons to $10^3$ TeV or higher. 
\end{abstract}

\maketitle

\section{Introduction}
\label{sec:intro}
The six orders of magnitude separation observed between the masses of the elementary quarks and the charged leptons adequately support the possibility that some of these fermion masses are induced radiatively. In the simplest version of this idea, only the third-generation fermions are postulated to have tree-level masses. At the same time, those of the first and second generations obtain their masses through higher-order corrections in a perturbation theory. It was realised from the very beginning through the early attempts \cite{Georgi:1972hy,Mohapatra:1974wk,Barr:1978rv,Wilczek:1978xi,Yanagida:1979gs,Barbieri:1980tz} that successful implementation of such an idea necessarily requires the extension of the Standard Model (SM). With the nonzero Yukawa couplings  for only the third-generation fermions at the tree level, the SM has a global $[U(2)]^5$ symmetry which remains unbroken by the electroweak symmetry breaking and prevents the corresponding two generations of fermions from getting mass through radiative corrections. Therefore, extensions of the SM in which the new sector breaks at least partially this global symmetry are desired in order to give rise to non-vanishing loop-induced masses for the first and second families of fermions.

The extended models for radiative fermion masses can mainly be categorized into two classes based on the nature of new particle(s) propagating in the  fermion self-energy loops\footnote{It is possible that the extensions include the scalar and vector bosons, and both can give rise to loop-induced corrections. However, the model can be put into one of these categories depending on the most-dominant contribution considered.} (a) spin-0 (see \cite{Balakrishna:1987qd,Balakrishna:1988ks,Balakrishna:1988xg,Balakrishna:1988bn,Babu:1988fn,Babu:1989tv,Arkani-Hamed:1996kxn,Barr:2007ma,Graham:2009gr,Dobrescu:2008sz,Crivellin:2010ty,Crivellin:2011sj,Chiang:2021pma,Chiang:2022axu,Baker:2020vkh,Baker:2021yli,Yin:2021yqy} for example) and (b) spin-1 \cite{HernandezGaleana:2004cm,Appelquist:2006ag,Reig:2018ocz,Weinberg:2020zba,Jana:2021tlx,Mohanta:2022seo}. In the case of the latter, SM gauge symmetry needs to be extended to include a gauged flavour symmetry group $G_F$ and the radiative correction can be determined in terms of the  gauge couplings and masses of the new gauge bosons. Since these couplings and masses are independently measurable parameters, the quantum-corrected fermion masses become in-principle  calculable quantities in this class of frameworks distinguishing them from those in category (a). Moreover, symmetry-based extensions of this kind typically lead to fewer parameters than extensions with scalars and hence can provide more predictive setups. Therefore, the radiative models in which the new gauge sector primarily induces the masses of the lighter families of the quarks and leptons can provide potentially more attractive and economical options and require systematic investigations.

Along these lines, we have recently proposed a model for $G_F$ being an abelian group and analysed it in detail in \cite{Mohanta:2022seo}. It was shown that the nature of the new abelian symmetry must be flavour non-universal if the fermion masses are to be generated radiatively. The simplest viable and anomaly-free implementation of the scheme requires $G_F = U(1)_1 \times U(1)_2$ where the gauge bosons associated with $U(1)_1$ and $U(1)_2$ generate the masses of the second and first generations at 1-loop, respectively. The inter-generational hierarchy between the first two generations can arise either from the hierarchy between the gauge boson masses or the difference between the strength of the gauge couplings of the two $U(1)$s. Some of these  assumptions can be alienated and a more predictive setup can be achieved if $U(1)_1 \times U(1)_2$ is replaced by a simple group. The most advantageous feature of non-abelian $G_F$ in the context of radiative mass generation is that it naturally accommodates gauge bosons with flavour non-diagonal couplings. The same symmetry can also be effectively utilized in order to ensure that only the third generations receive mass at the tree level. Moreover, being a simple group it minimally modifies the SM gauge structure and can lead to a predictive scenario.

In the present work, we investigate a scheme based on $G_F = SU(3)_F$ and provide a concrete and realistic implementation of this scheme for radiative mass induction for the lighter generations of the SM fermions. The horizontal $SU(3)$ symmetry was also proposed earlier in \cite{HernandezGaleana:2004cm,Hernandez-Galeana:2015zap} for a similar purpose, however systematic and comprehensive analysis of loop-induced fermion masses and mixing parameters along with the phenomenological constraints on the flavour symmetry breaking scale have not been carried out. Another non-abelian alternative, namely $G_F = SO(3)_L \times SO(3)_R$, has been investigated relatively recently in \cite{Weinberg:2020zba} and shown to lead to an inconsistent flavour spectrum. The present work, therefore, offers a complete and realistic model of radiatively induced quark and lepton masses based on non-abelian flavour symmetry. We find that a viable implementation of this scheme within the SM requires a multiplicity of the electroweak Higgs doublets  and the existence of vectorlike fermions. The latter plays an essential role in reproducing the observed spectrum being consistent with the constraints from the flavour violation. We show that the hierarchy between the first and second-generation masses can naturally be induced if the flavour symmetry is broken in a particular way.  This along with improved predictivity makes the present model less ad-hoc than the one based on abelian symmetries discussed earlier.

The rest of the paper is structured as the following. In the next section, we discuss the general framework of $SU(3)_F$ and the generation of radiative masses. Breaking of the horizontal symmetry leading to desired gauge boson mass spectrum is presented in section \ref{sec:GBmass}. Detailed implementation of the general scheme in the SM is discussed in section \ref{sec:model}. Viability of the model is established through example numerical solutions in section \ref{sec:numerical}. We also discuss some phenomenological aspects of the scheme in section \ref{sec:fv} before concluding in section \ref{sec:concl}.

\section{$SU(3)_F$ and fermion mass generation}
\label{sec:su3fm}
Denoting the three generations $(i=1,2,3)$ of chiral fermions by $f_{L i}^\prime$ and $f_{R i}^\prime$ and a pair of vectorlike fermions by $F^\prime_{L,R}$, the tree-level mass term in the basis $f^\prime_{ L \alpha} \equiv (f^\prime_{L i}, F^\prime_L)$ and $f^\prime_{R \alpha} \equiv (f^\prime_{R i}, F^\prime_R)$ is arranged as
\be \label{L_mass}
-{\cal L}_{m} = \overline{f}^\prime_{L \alpha}\, \left({\cal M}^0\right)_{\alpha \beta}\, f^\prime_{R \beta} + {\rm h.c.}\,, \ee
with
\be \label{Mf0}
{\cal M}^0 = \left( \ba{cc} 0_{3 \times 3} & (\mu)_{3 \times 1} \\ (\mu^\prime)_{1 \times 3} & m_F \ea \right)\,.\ee
Here, $f^\prime_{L,R}$ is used to discuss the general case in this section and $f=u,d,e$ will be used later to apply this discussion to the up-type, down-type quarks and charged leptons, respectively. For the brevity, we also suppress the $f$ dependency in ${\cal M}^0$, $\mu$ and $\mu^\prime$.

We also consider that $f^\prime_{L i}$ and $f^\prime_{R i}$ transform as fundamental representations of a horizontal gauged symmetry $SU(3)_F$. At the same time, the vectorlike fermions are taken as singlets under the same symmetry. It can then be seen that $\mu$ and $\mu^\prime$ matrices in the above mass Lagrangian break the $SU(3)_F$. The vanishing $3\times 3$ sub-matrix can be obtained utilizing the chiral nature of $f^\prime_{L i}$ and $f^\prime_{R i}$ under the SM gauge symmetry and it can remain zero even in the broken phase of $SU(3)_F$. Depending on the SM charges of $F^\prime_L$ or $F^\prime_R$, either  $\mu$ or $\mu^\prime$ is also protected by the chiral symmetry.

The matrix ${\cal M}^0 $ leads to two massless states which can be identified with the first and second-generation fermions. For $m_F \gg \mu_i, \mu^\prime_i$, the effective $3\times3$ mass matrix takes the form
\be \label{M_eff}
M^{0} \simeq -\frac{1}{m_F}\, \mu \,\mu^{\prime }\,, \ee
at the leading order. The above matrix is of rank one and the state corresponding to the non-vanishing eigenvalue can be assigned to the third generation fermion.

The relatively small masses of the first two generations can be induced through quantum corrections within this framework. In order to quantify these corrections, consider the $SU(3)_F$ gauge interactions of fermions with the gauge bosons $A_\mu^a$ given by
\be \label{L_gauge}
-{\cal L}_{\rm gauge} = g_F  \left( \overline{f}^\prime_{L i} \gamma^\mu A^a_\mu \left(\frac{\lambda^a}{2}\right)_{ij}f^\prime_{L j } +  \overline{f}^\prime_{R i} \gamma^\mu A^a_\mu \left(\frac{\lambda^a}{2}\right)_{ij}f^\prime_{R j } \right)\,, \ee
where $a =1,..,8$ and $\lambda^a$ are the Gell-Mann matrices. For the latter, we use the expressions in a different basis than the conventional one and they are listed in Appendix \ref{app:GM} for the clarity. The above can be generalized to include the vectorlike fermions as
\be \label{L_gauge_2}
-{\cal L}_{\rm gauge} = \frac{g_F}{2}  \left( \overline{f}^\prime_{L \alpha} \gamma^\mu A^a_\mu \left( \Lambda^a \right)_{\alpha \beta}f^\prime_{L \beta} +  \overline{f}^\prime_{R \alpha} \gamma^\mu A^a_\mu \left( \Lambda^a \right)_{\alpha \beta}f^\prime_{R \beta}  \right)\,, \ee
where $\Lambda^a$ are $4 \times 4$ matrices given by
\be \label{Lambda}
{ \Lambda}^a= \, \left(\ba{cc}\lambda^a& 0\\0 &0\ea \right)\,.\ee

The physical basis of fermions, denoted by $f_{L,R}$, can be obtained from the canonical basis using the unitary transformations
\be \label{trs}
f^\prime_{L,R}= {\cal U}_{L,R}\, f_{L,R}\,, \ee
such that 
\be \label{diag}
{\cal U}^{ \dagger}_L\, {\cal M}^{0}\, {\cal U}_R = {\cal D} = {\rm Diag.}(0,0,m_3,m_4)\,. \ee
Similarly, the physical gauge bosons  $B_\mu^a$ can be obtained from $A_\mu^a$ using an $8 \times 8$ real orthogonal matrix ${\cal R}$ as
\be \label{AtoB}
A_{a \mu}\,=\, {\cal R}_{ab}\, B_{b \mu} \,.\ee
The matrix ${\cal R}$ can be obtained explicitly by diagonalizing the gauge-boson mass matrix which is real and symmetric.

Substituting Eqs. (\ref{trs},\ref{AtoB}) in Eq. (\ref{L_gauge_2}), the gauge interactions in the physical basis of fermions and gauge bosons are obtained as
\be \label{L_gauge_3}
-{\cal L}_{\rm gauge} = \frac{g_F}{2}  \left(\overline{f}_{L \alpha} \gamma^\mu  \left({\cal U}_L^\dagger{\Lambda^a} {\cal U}_L \right)_{\alpha \beta}f_{L \beta} + \overline{f}_{R \alpha} \gamma^\mu  \left({\cal U}_R{}^\dagger{\Lambda^a} {\cal U}_R \right)_{\alpha \beta}f_{R \beta} \right) {\cal R}_{ab}B^b_\mu\,. \ee
Since $\Lambda^a$ do not commute with each other for all $a$, the matrices ${\cal U}_{L,R}^\dagger{\Lambda^a} {\cal U}_{L,R}$ cannot be made simultaneously diagonal. Therefore, there always exists a set of gauge bosons which has flavour-changing interactions with fermions. As noted by us previously in \cite{Mohanta:2022seo}, this is necessary for the generation of masses for the first and second family fermions through radiative corrections induced by the gauge bosons.

The fermion mass matrix, corrected by the $SU(3)_F$ gauge interactions at 1-loop, can be written as
\be \label{1loopmass}
{\cal M}= {\cal M}^{0}+ \delta {\cal M}\,, \ee 
where 
\be \label{dm}
\delta {\cal M} = {\cal U}_{L}\, \Sigma(0)\, {\cal U}^{ \dagger}_{R}\, .\ee
\begin{figure}[!t]
    \centering
    \includegraphics[width=8cm]{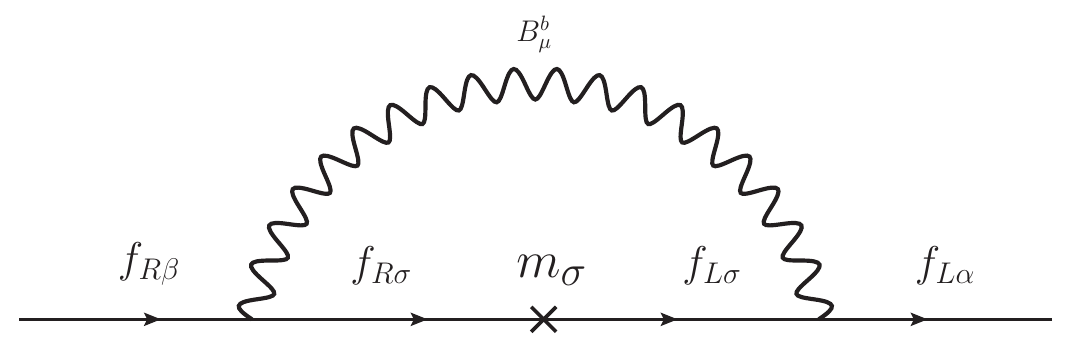}
    \caption{Diagram representing gauge boson induced fermion self-energy correction.}
    \label{fig:loop}
\end{figure}
Using Eq. (\ref{L_gauge_3}), the 1-loop contribution (see Fig. \ref{fig:loop}) can be computed as 
\beqa \label{sigma_1}
-i(\Sigma(p))_{\alpha \beta} &=& \int \frac{d^4 k}{(2\pi)^4}\left(-i\frac{g_F}{2} {\cal R}_{ab} ({\cal U}_L^{ \dagger} {\Lambda^a} {\cal U}_L)_{\alpha \sigma}\right) \gamma^\mu \frac{i m_{\sigma}}{ (k+p)^2-m_{\sigma}^2 + i\epsilon} \, \nonumber \\
& & \left(-i\frac{g_F}{2} {\cal R}_{c b} ({\cal U}_R^{ \dagger}{\Lambda^c}{\cal U}_R)_{\sigma \beta}\right) \gamma^\nu\, \Delta_{\mu \nu}(k)\,,\eeqa
with 
\be \label{propa}
\Delta_{\mu \nu}(k) = \frac{-i}{k^2-{M_{b}}^2 + i\epsilon} \left( \eta_{\mu \nu}-(1-\zeta)\frac{k_\mu k_\nu}{k^2-\zeta M_{b}^2}\right)\,.\ee
Here $M_{b}$ is the  mass of  the  gauge boson $B^b_{\mu}$. In the Feynmann-'t Hooft gauge, the evaluation of the above integral results in
\be  \label{sigma_2}
(\Sigma_f(0))_{\alpha \beta} = \frac{{g_F}^2}{16\pi^2}\, {\cal R}_{ab} ({\cal U}_L^\dagger{\Lambda^a}{\cal U}_L)_{\alpha \sigma} 
{\cal R}_{c b}({\cal U}_R^\dagger{\Lambda^c}{\cal U}_R)_{\sigma \beta}\, m_{\sigma}\, B_0[M_{b}^2,m_{\sigma}^2]\,, \ee 
with 
\beqa
B_0[M^2_{},m^2] &=& \frac{({2 \pi \mu})^\epsilon}{i \pi^2} \int d^dk\, \frac{ 1 }{k^2-m_{}^2 + i\epsilon}\frac{1}{k^2-{M_{}}^2 + i\epsilon} \, \nonumber \\
& = & \Delta_\epsilon -  \frac{M^2 \ln M^2 - m ^2 \ln m^2}{M^2 - m ^2}\,, \eeqa
and \be \Delta_\epsilon \,=\,\frac{2}{\epsilon}+1-\gamma + \ln{4\pi}\,. \ee

 It is straightforward to verify from Eq. (\ref{sigma_2}) that the $m_\sigma$ independent contribution in $\delta {\cal M}$ from $B_0[M_b^2,m_\sigma^2]$ vanishes identically. The divergent part of $\delta {\cal M}$ is given by
\be
\delta {\cal M}_{\rm div} = {\cal U}_{L}\, \Sigma(0)_{\rm div}\, {\cal U}^{ \dagger}_{R}\,,\ee
where $\Sigma(0)_{\rm div}$ captures the terms dependent only on $\Delta_\epsilon$ from Eq. (\ref{sigma_2}). Explicitly,
\beqa 
(\delta {\cal M}_{\rm div})_{\rho \kappa}&=&  
  \frac{{g_F}^2}{16\pi^2}\,{\cal U}_{L}{}_{\rho \alpha} {\cal R}_{ab} ({\cal U}_L^\dagger{\Lambda^a}{\cal U}_L)_{\alpha \sigma} 
{\cal R}_{c b}({\cal U}_R^\dagger{\Lambda^c}{\cal U}_R)_{\sigma \beta}\, m_{\sigma}\, \Delta_\epsilon\, {\cal U}^{ \dagger}_{R}{}_{\beta \kappa}\, , \nonumber\\
&=& \frac{{g_F}^2 \Delta_\epsilon}{16\pi^2}\, ({\cal R R^T})_{ac} \,{\cal U}_{L}{}_{\rho \alpha}\, ({\cal U}_L^\dagger{\Lambda^a}{\cal U}_L)_{\alpha \sigma}\, {\cal D}_{\sigma \sigma}\,({\cal U}_R^\dagger{\Lambda^c}{\cal U}_R)_{\sigma \beta}\,{\cal U}^{ \dagger}_{R}{}_{\beta \kappa}\,. \eeqa 
Using Eq. (\ref{diag}), orthogonality of ${\cal R}$ and unitarity of ${\cal U}_{L,R}$, the above expression can be simplified to 
\be \label{DM_div}
\delta {\cal M}_{\rm div} = \frac{{g_F}^2 \Delta_\epsilon}{16\pi^2}\, \Lambda^a {\cal M}^{0}\Lambda^a = 0\,,
\ee
where we use the form of ${\cal M}^{0}$ and $\Lambda^a$ given in Eqs. (\ref{Mf0},\ref{Lambda}) to get the last equality. The vanishing of $\delta {\cal M}_{\rm div}$ is in accordance with the renormalizability \cite{Barr:1978rv,Weinberg:1972ws} of the theory as there are no corresponding counterterms to renormalise.

Further simplification of the finite contribution can be achieved for $m_F \gg \mu_i, \mu^\prime_i$. In this case, ${\cal U}_{L,R}$ at the leading order in $\mu_i/m_F$ and $\mu^\prime_i/m_F$ can be approximated as \cite{Joshipura:2019qxz}
\be \label{U_block}
{\cal U}_{L,R} = \left(\ba{cc} U_{L,R} & -\rho_{L,R} \\ \rho_{L,R}^\dagger U_{L,R} & 1\ea\right) + {\cal O}(\rho_{L,R}^2)\,,\ee
where $\rho_L = -m_F^{-1}\mu$ and $\rho_R{}^\dagger = -m_F^{-1}\mu^\prime$. $U_{L,R}$ are $3 \times 3$ matrices that diagonalize $M^{0}$ given in Eq. (\ref{M_eff}) such that
\be \label{Meff_diag}
U_L^\dagger\,M^{0}\,U_R = {\rm Diag.}(0,0,m_3)\,.\ee
Using Eq. (\ref{M_eff}), the above can also be written as
\be \label{Meff_diag_comp}
\left(U_L\right)_{i3} \left(U_R^*\right)_{j3}\, m_3\, =M^0_{ij} = -\frac{1}{m_F}\mu_i \mu^\prime_j\,.\ee

Expanding the finite part of Eq. (\ref{sigma_2}) and using Eqs. (\ref{Lambda},\ref{U_block},\ref{Meff_diag_comp}), the 1-loop correction to the effective $3 \times 3$ mass matrix can be simplified to 
\be \label{dm_ij}
(\delta M)_{ij} \simeq \frac{{g_F}^2}{16 \pi^2} {\cal R}_{ab} {\cal R}_{cb} (\lambda^a M^0\lambda^c)_{ij}\, \Delta b_0[M_b^2]\,, \ee
where
\be \label{delta_b0} 
\Delta b_0[M_b^2]  = -  \frac{M_b^2 \ln M_b^2 - m_3^2 \ln m_3^2}{M_b^2 - m_3^2} + \frac{M_b^2 \ln M_b^2 - m_4^2 \ln m_4^2}{M_b^2 - m_4^2} \, .\ee
We also find $\delta{\cal M}_{4 \alpha} = \delta{\cal M}_{\alpha 4}=0$ by utilising Eq. (\ref{Lambda}). The non-observations of new gauge-bosons or vectorlike states would imply $m_3 \ll M_b,m_4$. In this limit, the loop function in Eq. (\ref{dm_ij}) can be approximated to
\be \label{loop_approx}
\Delta b_0[M_b^2]  \simeq \frac{m_4^2}{M_b^2 - m_4^2}\, \ln\left(\frac{M_b^2}{m_4^2} \right)\,. \ee\

The following noteworthy features of the loop-corrected fermion mass matrix can be deduced from Eq. (\ref{dm_ij}) along with Eq. (\ref{loop_approx}) and the explicit expressions of the $\lambda^a$ given in Appendinx \ref{app:GM}.
\begin{itemize}
\item The loop-induced masses are suppressed by the loop factor $g_F^2/(16 \pi^2)$ if $m_4 > M_b$. An additional suppression by factor $m_4^2/M_b^2$ arises in case $m_4 < M_b$. 
\item For a generic choice of gauge boson masses and the orthogonal matrix ${\cal R}$, Eq. (\ref{dm_ij}) induces masses of the same order for both the first and second-generation fermions.
\item A phenomenologically desired possibility would be that only the second generation fermions become massive at 1-loop while the first generation remains massless and receive mass at higher order. Inspecting the expression Eq. (\ref{dm_ij}) and the Gell-mann matrices, we do not find a possibility in which the first-generation fermions can be made strictly massless at 1-loop.
\end{itemize}
The above results indicate that while the loop suppressed masses for the first and second generations masses naturally emerge, the hierarchy between the two requires additional arrangements. Utilizing the first feature mentioned above, we discuss a scenario in the next section which can lead to such a hierarchy within this framework.  
 
\section{ Gauge-boson mass hierarchy}
\label{sec:GBmass}
Consider a two-step breaking of $SU(3)_F$ symmetry such that 
\be \label{}
SU(3)_F\,  \xrightarrow{\langle \eta_1 \rangle}\, SU(2)_F\, \xrightarrow{\langle \eta_2 \rangle}\, {\rm nothing}\,,\ee
with $\langle \eta_1 \rangle \gg \langle \eta_2 \rangle$. With this arrangement, three of the gauge bosons corresponding to the generators of $SU(2)_F$ are expected to be lighter than the remaining five gauge bosons. This hierarchy among the gauge bosons ultimately results in the mass hierarchy between the first and second-generation fermions as we show below.

In the choice of our basis of Gell-Mann matrices, it is convenient to identify the intermediate $SU(2)_F$ with the generators $\lambda^\alpha$ with $\alpha = 1,2,3$. Denoting the remaining indices with $m=4,...8$, the gauge boson mass term in the canonical basis $A^\mu_a = \left(A^\mu_\alpha, A^\mu_{m} \right)$ can be written as
\be \label{LGB}
-{\cal L}^{M}_{\rm GB} = \frac{1}{2} {\cal M}^2_{a b}\,A^\mu_a A_{b \mu}\,, \ee
such that
\be \label{MGB_block}
{\cal M}^2 = \left( \ba{cc} M^2_{(33)} & M^2_{(35)}\\ (M^2_{(35)})^T & M^2_{(55)} \ea \right)\,.\ee
Here, $M^2_{(AB)}$  are $A \times B$ dimensional matrices sub-blocks of the gauge boson mass matrix. The two step breaking of the horizontal symmetry implies $M^2_{(33)}, M^2_{(35)} \ll M^2_{(55)}$. The gauge-boson mass matrix in this specific form can be diagonalized by see-saw like diagonalization procedure and the orthogonal matrix at the leading order can be written as 
\be \label{R_block}
{\cal R} = \left(\ba{cc} R_{3} & -\rho R_5 \\ \rho^T R_3 & R_5 \ea\right) + {\cal O}(\rho^2)\,, \ee
with $\rho = -M^2_{(35)} (M_{(55)}^2)^{-1}$. Here, $R_3$ and $R_5$ are real orthogonal matrices of dimensions $3\times3$ and $5\times5$, respectively.

Substituting Eq. (\ref{R_block}) in Eq. (\ref{dm_ij}) and considering the leading order terms in the seesaw expansion parameter $\rho$, we get
\beqa \label{dmij2}
({\delta M})_{ij} = \frac{{g_F}^2}{16 \pi^2} &\Big[& (R_3)_{\alpha \beta}(R_3)_{\gamma \beta} \left(\lambda^\alpha M^0 \lambda^\gamma \right)_{ij} \Delta b_0[M_\beta^2] \Big. \nonumber \\
& + & (R_3)_{\alpha \beta}(\rho^T R_3)_{m\beta} \left(\lambda^\alpha M^0\lambda^m + \lambda^m M^0\lambda^\alpha \right)_{ij}  \Delta b_0[M_\beta^2] \nonumber \\
& + & (R_5)_{mn}(R_5)_{pn} \left(\lambda^m M^0\lambda^p \right)_{ij} \Delta b_0[M_n^2] \nonumber \\
& - & \Big. (R_5)_{mn}(\rho R_5)_{\alpha n} \left(\lambda^m M^0\lambda^\alpha + \lambda^\alpha M^0\lambda^m \right)_{ij}  \Delta b_0[M_n^2]\, +\, {\cal O}(\rho^2) \Big]\,. \eeqa
Recall that $\alpha,\beta,... = 1,2,3$ while $m,n,... = 4,...,8$. The first term gives the dominant contribution to $\delta M$ as $M_\alpha^2 < M_m^2$. Since $\lambda^\alpha$ has a vanishing first row and first column, this contribution is of rank one and it induces only the mass for the second generation fermion. For $M_\alpha < m_4$, this mass is suppressed by only the loop factor in comparison to that of the third generation. The masses of the first-generation fermions arise from the remaining terms in Eq. (\ref{dmij2}) and it is suppressed either by $M_\alpha^2/M_m^2$ or $m_4^2/M_m^2$ with respect to the second-generation mass. In this way, the 1-loop induced corrections can give rise to the desired hierarchy between the fermion masses if 
\be \label{scales}
M_\alpha^2  \lesssim m_4^2 \lesssim M_m^2\,. \ee
This implies that the scale of $SU(3)_F$ breaking and the mass scale of vectorlike fermions are required to be close to each other. However, the overall scale of these new states is not constrained from the pure consideration of fermion masses as the finite corrections always come as a ratio of the $m_4$ and $M_a$.

\section{Model implementation}
\label{sec:model}
Based on the general conditions to generate fermion mass hierarchy through quantum corrections, we now give a specific and minimal implementation of the general framework in which all these aspects can be realised explicitly. As anticipated, we consider three generations of SM fermions transforming as fundamental representations of the horizontal gauged symmetry $SU(3)_F$. We additionally consider $N_R$, triplet of three SM singlet fermions under $SU(3)_F$, which is necessary for anomally cancellation. The SM Higgs is replaced by the two Higgs doublets, each of them comes also in three copies to form triplets of $SU(3)_F$. Two additional SM singlet and $SU(3)_F$ triplets scalars, $\eta_s$, are introduced for the consistent gauge symmetry breaking and also to give rise to the desired fermion mass matrices at the tree level. As already outlined in section \ref{sec:su3fm}, the framework requires vectorlike fermions which are assumed singlets under the new symmetry. The matter and scalar fields along with their SM and $SU(3)_F$ transformation properties are summarised in Table \ref{tab:fields}.
\begin{table}[!t]
\begin{center}
\begin{tabular}{ccc} 
\hline
\hline
~~Fields~~&~~$(SU(3)_c \times SU(2)_L \times U(1)_Y)$~~&~~$SU(3)_F$~~~~\\
\hline
$Q_{L} $ & $(3,2,\frac{1}{6}) $ & 3 \\
$u_{R} $ & $(3,1,\frac{2}{3}) $ & 3 \\
$d_{R} $ & $(3,1,-\frac{1}{3}) $ & 3\\
\hline
$L_{L} $ & $(1,2,-\frac{1}{2}) $ & 3 \\
$e_{R} $ & $(1,1,-1) $ & 3 \\
$N_{R} $ & $(1,1,0) $ & 3 \\
\hline
$H_{u}$ & $(1,2,-\frac{1}{2}) $ & 3 \\
$H_{d}$ & $(1,2,\frac{1}{2}) $ & 3 \\
{$\eta_1$,\,$\eta_2$}  &{ $(1,1,0) $} & ${3}$ \\
\hline
$T_L, \, T_R $ & $(3,1,\frac{2}{3}) $ & 1\\
$B_L, \, B_R $ & $(3,1,-\frac{1}{3}) $ & 1\\
$E_L, \, E_R $ & $(1,1,-1) $ & 1\\
\hline
\end{tabular}
\end{center}
\caption{The SM and $G_F$ charges of various fermions and scalars of the model.}
\label{tab:fields}
\end{table}

\subsection{$SU(3)_F$ breaking and gauge boson mass ordering}
\label{subsec:su3breaking}
The non-observations of $SU(3)_F$ gauge bosons in the experiments so far imply that the scale of the latter's  breaking is much larger than the weak scale. Therefore, the new gauge symmetry must be primarily broken by the SM singlet fields $\eta_{1,2}$ and the contributions from the electroweak doublets are expected to be sub-dominant. With this reasoning, we consider the breaking of $SU(3)_F$ driven by $\eta_{1,2}$ only.

As an example, we consider the following VEV configuration:
\be \label{eta_vevs}
\langle \eta_1 \rangle = (v_F,0,0)^T\,,~~\langle \eta_2 \rangle = (0,0,\epsilon v_F)^T\,,\ee
with $\epsilon < 1$. One of these can always be chosen in the given form using the $SU(3)_F$ rotation without losing generality. Therefore, the single field does not break the gauge symmetry completely and leaves its  $SU(2)$ subgroup unbroken. This requires at least two scalars to break fully the $SU(3)_F$ symmetry with VEVs in different directions. For simplicity, we choose the other VEVs in a specific direction orthogonal to the first one. We write down the most general gauge invariant potential involving $\eta_{1,2}$  in Appendinx \ref{app:potential} and show that the VEV configuration given in Eq. (\ref{eta_vevs}) can be obtained by a suitable choice of parameters in the scalar potential.

The kinetic terms of $\eta_{1,2}$ after the spontaneous breaking of $SU(3)_F$ leads to the following gauge boson mass matrix defined in Eq. (\ref{LGB}):
\be \label{MGB}
{\cal M}^2_{\alpha \beta} =\frac{g_F^2}{2}\,\sum_{s=1,2} \langle\eta_s\rangle^\dagger \lambda^{\alpha \dagger} \lambda^{\beta} \langle\eta_s\rangle\,.\ee
In the notation of Eq. (\ref{MGB_block}), the above mass matrix can be written as 
\beqa \label{MGB_block2}
M^2_{(33)} &=& \frac{g_F^2 v_F^2}{2}\,{\rm Diag.}\left(\epsilon^2, \epsilon^2, \epsilon^2\right)\,, \nonumber \\
M^2_{(55)} &=& \frac{g_F^2 v_F^2}{2}\,{\rm Diag.}\left(1, 1, 1+\epsilon^2, 1+\epsilon^2, \frac{1}{3}(4 + \epsilon^2) \right)\,, \nonumber \\
M^2_{(35)} &=& \frac{g_F^2 v_F^2}{2}\,\left(\ba{ccccc} 0 & 0 & 0 & 0 & 0 \\
0 & 0 & 0 & 0 & 0 \\
0 & 0 & 0 & 0 & -\frac{\epsilon^2}{\sqrt{3}} \\  \ea \right) \eeqa
The structure of the gauge boson mass matrix in this case is extremely simple and there exists mixing between only $A^\mu_3$ and $A^\mu_8$ states which correspond to diagonal generators.

The matrix ${\cal M}^2$ can be diagonalized by ${\cal R}$ as parametrized by Eq. (\ref{R_block}) with the following explicit forms of its sub-matrices:
\be \label{R_res}
R_3 = \mathbb{I}_{3 \times 3}\,,~~R_5 = \mathbb{I}_{5 \times 5}\,,~~\rho = \left(\ba{ccccc} 0 & 0 & 0 & 0 & 0 \\
0 & 0 & 0 & 0 & 0 \\
0 & 0 & 0 & 0 & -\frac{\sqrt{3}}{4}\epsilon^2 \\  \ea \right)\,.\ee
The diagonal gauge boson mass matrix is then obtained as
\be \label{MGB_diag}
{\cal D}^2 = \frac{g_F^2 v_F^2}{2}\,{\rm Diag.}\left(\epsilon^2, \epsilon^2, \epsilon^2, 1, 1, 1+\epsilon^2, 1+\epsilon^2, \frac{4}{3}+\frac{1}{3}\epsilon^2 \right)\, +\,  {\cal O}(\epsilon^4)\,. \ee
In this setup, one obtains the hierarchical gauge boson mass spectrum $M^2_{1,2,3} \ll M^2_{4,...,8}$ as required to generate the mass gaps between the first and second-generation fermions.

Substituting the above in the $(\delta M)_{ij}$, we find
\beqa \label{dmij3}
\frac{16 \pi^2}{g_F^2}(\delta M)_{ij} &= & \sum_{\alpha=1}^3 \left(\lambda^\alpha M^{0} \lambda^\alpha \right)_{ij} \Delta b_0 [M_\alpha^2] + \sum_{m=4}^8 \left(\lambda^m M^{0} \lambda^m \right)_{ij} \Delta b_0[M_m^2]  \nonumber \\
& + & \frac{\sqrt{3}}{4}\epsilon^2 \left(\lambda^3 M^{0} \lambda^8 + \lambda^8 M^{0} \lambda^3 \right)_{ij}  \left(\Delta b_0[M_3^2] - \Delta b_0[M_8^2] \right)\,+\,{\cal O}(\epsilon^4)\,. \eeqa
An approximate degeneracy of some of the gauge bosons allows further simplification. Setting
\be \label{MZ1-MZ2}
M_1^2 \simeq M_2^2 \simeq M_3^2 \equiv M_{Z_1}^2\,,~~ M_4^2 \simeq ... \simeq M_7^2 \simeq \frac{3}{4} M_8^2 \equiv M_{Z_2}^2\,,\ee
and using $\epsilon^2 = M_{Z_1}^2/M_{Z_2}^2$, we find
\beqa \label{dmij4}
\frac{16 \pi^2}{g_F^2}\,{\delta M} & \simeq & \left(\ba{ccc} 0 & 0 & 0\\ 0 & M^0_{22} + 2 M^0_{33} & -M^0_{23} \\ 0 & - M^0_{32} & 2 M^0_{22} + M^0_{33} \ea \right)\, \Delta b_0[M_{Z_1}^2]\nonumber \\
& + &  \left(\ba{ccc} 2(M^0_{22} + M^0_{33}) & 0 & 0\\ 0 & 2 M^0_{11}  & 0 \\ 0 & 0 & 2 M^0_{11}  \ea \right)\, \Delta b_0[M_{Z_2}^2]\nonumber \\
& + & \frac{1}{3} \left(\ba{ccc} 4 M^0_{11}   & -2 M^0_{12}  &  -2 M^0_{13} \\ -2 M^0_{21}  & M^0_{22}   &  M^0_{23}  \\ -2 M^0_{31}  & M^0_{32}  & M^0_{33}   \ea \right)\, \Delta b_0\Big[\frac{4}{3}M_{Z_2}^2\Big] \nonumber \\
& + & \frac{M_{Z_1}^2}{2 M_{Z_2}^2} \left(\ba{ccc} 0   & - M^0_{12}  &   M^0_{13} \\ - M^0_{21}  & M^0_{22}   & 0  \\  M^0_{31}  & 0  & -M^0_{33}   \ea \right) \left(\Delta b_0[M_{Z_1}^2] - \Delta b_0\Big[\frac{4}{3}M_{Z_2}^2\Big] \right). \eeqa

\subsection{Charged fermion masses}
\label{subsec:CFM}
With the set of fields and their transformation properties defined in Table \ref{tab:fields}, the most general renormalizable Yukawa Lagrangian  of the model can be written as 
\beqa \label{yukawa}
-{\cal L}_{Y}&=& {y}_u \, \overline{Q'}_{Li} H^i_u  T'_{R}\, +   {y}_d \, \overline{Q'}_{L i} H_d^i  B'_R\, + {y}_e \, \overline{L'_i}_L H_d^i  E^\prime_R\,  \nonumber \\
&+& {y}'^{(s)}_u \,\overline{T'}_{L} \eta^\dagger_{s i}\,  u'^{i}_{R} \,+ {y}'^{(s)}_d \,\overline{B'}_L \eta^\dagger_{s i}\,  d^{\prime i}_R\, +  {y}'^{(s)}_e \,\overline{E'}_L \eta^\dagger_{s i}\,  e^{\prime i}_R \, \nonumber \\
& + & m_T \overline{T'}_L T'_R+\, m_B \overline{B'}_L B'_R+\, m_E \overline{E^\prime}_L E^\prime_R\,+{\rm h.c.}\, \eeqa
where $ i=1,2,3$ is an $SU(3)_F$ index and $s= 1,2$ denotes multiplicity of $\eta$ fields. The primed notation is used for the fields in a flavour basis.

It is straightforward to see that after the electroweak and $SU(3)_F$ breaking, the Yukawa interactions in Eq. (\ref{yukawa}) lead to the tree-level mass matrices in the desired form of Eq. (\ref{Mf0}) with 
\be \label{mumup} 
\mu_f\,=\,\left(\ba{ccc}y_f v^f_1 & y_f v^f_2&y_f v^f_3 \ea \right)^T\,~ \text{and }\,~ \mu^\prime_f\,=\,\left(\ba{ccc} y^{\prime (1)}_f\, v_F & 0 &y^{\prime (2)}_f \, \epsilon v_F \ea \right)\,,\ee
where $f=u, d, e$ which accounts for three type of charged fermions. Also, $v^u_i = \langle H_u^i \rangle$, $v^d_i = v^e_i = \langle H_d^i \rangle$. The VEVs of $\eta_{1,2}$ are taken from Eq. (\ref{eta_vevs}). The tree-level effective mass matrix after integrating out the heavy vectorlike states is given by
\be \label{Meff_0}
M_{u,d,e}^{0} \equiv - \frac{1}{m_{T,B,E}}\, \mu_{u,d,e}\, \mu^\prime_{u,d,e}\,.\ee
The matrices $M_{f}^{0}$ has vanishing second column.

At 1-loop, the charged fermion mass matrices are given by
\be \label{Meff_1}
M_f = M_f^0 + \delta M_f\,,\ee  
and $\delta M_f$ can be obtained using the general expression, Eq. (\ref{dmij4}). Using the form of the tree-level mass matrices, we get
\beqa \label{dm:model}
\delta M_f & \simeq & \frac{g_F^2 N_f}{16 \pi^2} \left[\left(\ba{ccc} 0 & 0 & 0\\ 0 &  2 (M^0_f)_{33} & -(M^0_f)_{23} \\ 0 & 0 &   (M^0_f)_{33} \ea \right)\, \Delta b_0[M_{Z_1}^2] \right. \nonumber \\
& + & 2 \left(\ba{ccc} (M^0_f)_{33} & 0 & 0\\ 0 & (M^0_f)_{11}  & 0 \\ 0 & 0 & (M^0_f)_{11}  \ea \right)\, \Delta b_0[M_{Z_2}^2]\nonumber \\
& + & \frac{1}{3} \left(\ba{ccc} 4 (M^0_f)_{11}   & 0  &  -2 (M^0_f)_{13} \\ -2 (M^0_f)_{21}  & 0   &  (M^0_f)_{23}  \\ -2 (M^0_f)_{31}  & 0  & (M^0_f)_{33}   \ea \right)\, \Delta b_0\Big[\frac{4}{3}M_{Z_2}^2\Big] \nonumber \\
& + & \left. \frac{1}{2} \frac{M_{Z_1}^2}{M_{Z_2}^2} \left(\ba{ccc} 0   & 0  &   (M^0_f)_{13} \\ - (M^0_f)_{21}  & 0   & 0  \\  (M^0_f)_{31}  & 0  & -(M^0_f)_{33}   \ea \right)\, \left(\Delta b_0[M_{Z_1}^2] - \Delta b_0\Big[\frac{4}{3}M_{Z_2}^2\Big] \right) \right] \,, \eeqa
where $N_f = 3$ for $f=u,d$ and $N_f=1$ for $f=e$.  As we demonstrate in the next section, the above $M_f$ can reproduce the observed charged fermion mass spectrum and quark mixing.

\subsection{Neutrino masses}
\label{subsec:Neutrino}
As noted earlier, the anomaly-free model requires the SM singlet fermions $N_R$. With the present field content and the symmetry of the model, it is evident that there is no Dirac Yukawa coupling between $L_L$ and $N_R$ and no Majorana mass term for $N_R$ at the renormalizable level. This prevents $N_R$ from contributing to the light neutrino masses through the conventional type I seesaw mechanism.

 The symmetry of the model, however, allows the following Weinberg operators
\be \label{nu_op}
\frac{c_1}{\Lambda} \left(\overline{L_L^c}^i H_{u i}^* \right) \left(L_L^j  H_{u j}^\dagger \right) +\frac{c_2}{\Lambda} \left(\frac{\lambda^a}{2}\right)^k_i \left(\frac{\lambda^a}{2}\right)^l_j \left(\overline{L_L^c}^i  H_{u k}^* \right) \left(L_L^j   H_{u l}^\dagger \right) + \frac{c_3}{\Lambda} \left(\overline{L_L^c}^i L_L^j  \right) \left(H_{u i}^\dagger H_{u j}^* \right)\,,\ee
which can induce the suppressed neutrino masses in comparison to those of the charged fermions. It is straightforward to extend the model for an ultra-violet completion of the above operators. For instance, the simplest possibility is to introduce two or more fermions, namely $\nu_{R k}$, which are singlets under the full gauge symmetry and, therefore, do not give rise to anomalies. They can couple to $SU(3)_F$ triplet $L_L^i$ and anti-triplet $H_{u i}^\dagger$ giving rise to the usual Dirac Yukawa term and can also possess Majorana mass term unrestricted by the gauge  symmetry of the model. The first operator in Eq. (\ref{nu_op}) gets generated when $\nu_{R k}$ are integrated out from the spectrum. Similarly, the second and third operators can be induced by heavy $SU(3)_F$ adjoint fermions and sextet scalars respectively.

In contrast to charged fermions, neutrinos can acquire their masses at the tree level through dimension-5 operators without being restricted by the constraints of underlying flavour symmetries. This property is particularly advantageous because the inter-generational mass hierarchy among neutrinos is comparatively weaker than that observed among charged fermions. Consequently, the masses and mixing parameters of neutrinos remain relatively unconstrained within the framework of the effective theory. Nevertheless, it's worth noting that specific constraints could emerge depending on the chosen ultra-violet completion.

\section{Numerical Solutions}
\label{sec:numerical}
To establish the validity of the model and to understand the pattern of its various parameters, we carry out numerical analysis to find example solutions which can reproduce the observed values of the charged fermion masses and quark mixing parameters. Removing the unphysical phases through redefinitions of various matter fields in Eq. (\ref{yukawa}), it can be seen that the parameters $y_f$, $y_f^{\prime (1)}$ and $m_{T,B,E}$ can be made real. Moreover, we assume that all the VEVs are real. This leaves only three complex parameters, namely $y_{u,d,e}^{\prime (2)}$ in the model. Through Eq. (\ref{mumup}), this implies real $(\mu_f)_i$ (for $i=1,2,3$), real $(\mu_f^\prime)_1$ and a complex $(\mu_f^\prime)_3$. Moreover, 
\be \label{mue-mud}
\mu_e =\frac{y_e}{y_d} \mu_d \equiv r\, \mu_d\,,\ee
where $r$ is a real parameter. Altogether, there are $21$ real parameters (real $\mu_{ui}$, $\mu_{di}$, $r$, $\mu_{u1}^\prime$, $\mu_{d1}^\prime$, $\mu_{e1}^\prime$, $m_T$, $m_B$, $m_E$, $M_{Z_1}$, $M_{Z_2}$ and complex $\mu_{u3}^\prime$, $\mu_{d3}^\prime$, $\mu_{e3}^\prime$) in the model leading to $13$ observables (9 charged fermion masses, 3 quark mixing angles and a CP phase). Despite of a large number of parameters than the observables, it is not obvious that the model can viably reproduce the latter given various constraints and correlations among the parameters as we describe below. 

\begin{table}[t]
\begin{center}
\begin{tabular}{cccc} 
\hline
\hline
~~Parameters~~&~~Solution 1 (S1)~~&~~Solution 2 (S2)~~&~~Solution 3 (S3)~~\\
 \hline
$M_{Z_1}$ & $10^4$ GeV & $10^6$ GeV & $10^8$ GeV \\
$\epsilon$   & $ 0.0131 $ & $ 0.0083 $ & $ 0.0108 $ \\
$\epsilon_T$ & $ 0.3576 $  & $ 0.3455   $ & $ 0.4357 $ \\
$\epsilon_B$ & $ 0.9838  $ & $ 0.9999   $ & $0.9956  $ \\
$\epsilon_E$ & $ 0.8831  $ & $ 0.3881   $ & $ 0.7358 $ \\
\hline
$\epsilon_{u1}$ & $  0.4013  $  & $-0.3903  $ & $ -0.3745 $ \\
$\epsilon_{u2}$ & $  0.7211  $  & $ 0.6237  $ & $ -0.8653 $ \\
$\epsilon_{u3}$ & $ -0.8575 $   & $ 0.6714  $ & $ -0.8504 $ \\
\hline
$\epsilon^\prime_{u1}$ & $ 0.2970 $ & $0.3425   $ & $ -0.3393 $ \\
$\epsilon^\prime_{u3}$ & $ 0.0115  - i\, 0.32\times 10^{-4} $ & $ 0.0142  + i\,0.08\times 10^{-4}$ & $ 0.0140- i\,0.72\times 10^{-4} $ \\
\hline
$\epsilon_{d1}$ & $-0.1160  $ & $0.2935   $ & $ -0.2423  $ \\
$\epsilon_{d2}$ & $-0.2288  $ & $ -0.4345 $ & $ -0.5288  $ \\
$\epsilon_{d3}$ & $ 0.2494  $ & $ -0.5095 $ & $ -0.5668  $ \\
\hline
$\epsilon^\prime_{d1}$ & $ 0.0460  $ & $ 0.0227  $ & $-0.0205  $ \\
$\epsilon^\prime_{d3}$ & $ 0.0030  + i\,0.95 \times 10^{-3}$ & $ 0.0017 - i\,0.33\times 10^{-3}  $ & $ 0.0011+ i\,0.41 \times 10^{-3} $ \\
\hline
$r$ & $ 3.7895 $  & $-0.5749 $ & $-0.4040 $ \\
\hline
$\epsilon^\prime_{e1}$ & $-0.0023 $ & $-0.0013 $  & $-0.0079 $\\
$\epsilon^\prime_{e3}$ & $-0.0062 -i\, 0.28 \times 10^{-3}$ & $-0.0093-i\,0.86\times 10^{-3}  $ & $ 0.0214   + i\, 0.42 \times 10^{-3} $\\
\hline
\hline
\end{tabular}
\end{center}
\caption{Three benchmark solutions and the optimized values of the model parameters for different $M_{Z_1}$ which lead to viable charged fermion masses and quark mixing.}
\label{tab:sol}
\end{table}

For simplicity, various dimension-full parameters can be expressed in terms of the mass scales in the model and dimension-less quantities. Considering that viable fermion mass hierarchy would prefer $M_{Z_1}^2 \lesssim m_{T,B,E}^2 \lesssim M_{Z_2}^2$ (see Eq. (\ref{scales})), we define
\be \label{dimless1}
m_T = \epsilon_T M_{Z_2},\, m_B = \epsilon_B M_{Z_2},\, m_E = \epsilon_E M_{Z_2}. \ee
Also recall that $M_{Z_1} = \epsilon M_{Z_2}$. Moreover, we also define
\be \label{dimless2}
\mu_{f1 }^\prime = \epsilon_{f 1}^\prime M_{Z_2},\, \mu_{f3}^\prime = \epsilon_{f 3}^\prime M_{Z_2}\,,\ee
and 
\be \label{dimless3}
\mu_{ui} = \epsilon_{ui}  v,\, \mu_{di} = \epsilon_{di}  v, \ee
where $v=174$ GeV. In this way, various $\epsilon$ and $\epsilon^\prime$ can preferably take values less than unity.

Taking a particular value of $M_{Z_1}$, we obtain the values of the remaining dimensionless parameters using the $\chi^2$ optimization technique. Our methodology for the latter is described in detail in \cite{Mummidi:2021anm}. Three benchmark solutions obtained in this way are displayed in Table \ref{tab:sol} for different $M_{Z_1}$. The minimized values of $\chi^2$ are $ 6.97$, $6.90 $ and $6.46 $ for the solutions S1, S2 and S3, respectively. We also list the resulting values of charged fermion masses and quark mixing parameters for all three solutions in Table \ref{tab:res} along with the corresponding experimental values for  comparison.

\begin{table}[t]
\begin{center}
\begin{tabular}{ccccc} 
\hline
\hline
~Observable~& Value &~~S1~~&~~S2~~&~~S3~~\\
\hline
$m_u$ [MeV] & $1.27 \pm 0.5$    & $1.31$ &$1.26$     & $1.23$\\
$m_c$ [GeV] & $0.619 \pm 0.084$ &$0.567$  &$0.662$   & $0.612$\\
$m_t$ [GeV] & $171.7 \pm 3.0$   & $171.7$ &$171.6$   & $171.6$\\
\hline
$m_d$ [MeV] & $2.90 \pm 1.24$   & $3.71$  &$4.15$  & $3.55$\\
$m_s$ [GeV] & $0.055 \pm 0.016$ & $0.016$ &$0.018$  & $0.016$\\
$m_b$ [GeV] & $2.89 \pm 0.09$   & $2.89$  &$2.88$  & $2.89$\\
\hline
$m_e$ [MeV] & $0.487 \pm 0.049$    & $0.492$  &$0.487$   &$0.489$ \\
$m_\mu$ [GeV] & $0.1027 \pm 0.0103$   & $0.1007$  &$0.099$   & $0.1004$\\
$m_\tau$ [GeV] & $1.746 \pm 0.174$ & $1.784$  &$1.786$   & $1.777$\\
\hline 
$|V_{us}|$ & $0.22500 \pm 0.00067$ & $0.21614$  &$0.22242$  & $0.22226$\\
$|V_{cb}|$ & $0.04182 \pm 0.00085$ & $0.04110$  &$0.04270$  & $0.04207$\\
$|V_{ub}|$ & $0.00369 \pm 0.00011$ & $0.00363$  &$0.00378$  & $0.00371$\\
$J_{CP}$ & $(3.08 \pm 0.15)\times 10^{-5}$ & $3.14\times 10^{-5}$ & $3.02\times 10^{-5}$ & $3.06\times 10^{-5}$\\
\hline
\hline
\end{tabular}
\end{center}
\caption{The fitted values of the charged fermion masses and quark mixing parameters at the minimum of $\chi^2$ for three benchmark solutions are displayed in Table \ref{tab:sol}. The second column denotes experimentally measured value of corresponding observable extrapolated at $M_Z$ that has been used in the $\chi^2$ function.}
\label{tab:res}
\end{table}

Some of the noteworthy features of the model that can be derived from Table \ref{tab:sol} are as the following. We obtain almost similar values of the minimized $\chi^2$ for different values of $M_{Z_1}$. The ability to reproduce the realistic flavour hierarchies, therefore, depends on the relative masses of new gauge bosons and vectorlike states and not on the overall flavour symmetry breaking scale. This is expected as the flavour hierarchies are technically natural. All the $\epsilon_{fi}$ are of ${\cal O}(10^{-1})$ implying the fundamental Yukawa couplings $y_f$ of the same order and no large hierarchy between the VEVs, $v^{u,d}_i$. One also finds $\epsilon^\prime_{f3} < \epsilon^\prime_{f1}$ for $f=u,d$ as expected from Eq. (\ref{mumup}). The fitted values of $\epsilon^\prime_{e1}$, however, require two orders of separation between the magnitudes of $y_e^{\prime(1)}$ and $y_u^{\prime(1)}$. Altogether, the values of fundamental Yukawa couplings of the model range in just two orders of magnitude unlike in the SM where such a range spans at least five orders. Since all the third-generation fermions receive their masses at the tree level, the hierarchy between $m_t$ and $m_{b,\tau}$ does not follow naturally and requires $\epsilon_T \ll \epsilon_{B,E}$.

It can be  noticed from Table \ref{tab:res} that all the observables, except  $m_s$, are  fitted within $\pm  1\sigma$ range of their reference values for all the solutions.  The fitted value of $m_s$ is $\sim 2.4 \sigma$ away from the experimental value. Despite having a sufficiently large number of parameters than the observable, the inability to  reproduce the central value of $m_s$ indicates the existence of non-trivial correlations between the observables that result from the predictive nature of non-abelian flavour symmetry.  Remarkably, a more precise  measurement of strange quark mass can falsify the model irrespective of the scale of $SU(3)_F$  breaking.

\section{Flavour violation}
\label{sec:fv}
One of the most common features of radiative fermion mass models is the inherent presence of flavour-changing neutral currents. In the present framework, they arise from (i) flavour non-universal gauge interactions and (ii) mixing between the chiral and vectorlike fermions. Typically for $M_{Z_1} \ll m_{T,B,E}$, the first provides dominant contributions over the second and leads to a lower limit on the mass scales of the new fields. We study them in detail in this section by first deriving the general dimension-6 effective operators and then estimating various relevant quark and lepton flavour transitions.

Rewriting Eq. (\ref{L_gauge_2}) in the physical basis of fermions, one finds
\be \label{Lm_guage}
-{\cal L}_{\rm gauge} = \frac{g_F}{2}  \left( \overline{f}_{L i} \gamma^\mu   \left(\tilde{\lambda}^a_{f}{}_L \right)_{ij}  f_{L  j}\, +\,  \overline{f}_{R i} \gamma^\mu  \left(\tilde{\lambda}^a_{f}{}_R \right)_{ij}  f_{R j} \right)\, {\cal R}_{ab}B^b_\mu\,, \ee
where
\be \label{}
 \tilde{\lambda}^a_{f}{}_{L,R} = U^{f}{}^\dagger_{L,R}\, \lambda^a\, U^{f}{}_{L,R}\,,\ee
and $U^{f}{}_{L,R}$ are unitary matrices that diagonalize the 1-loop corrected $M_f$. Integrating out the gauge bosons, we find the effective dimension-6 operators as
\beqa \label{d6_ops}
{\cal L}_{\rm eff} =&& C^{(ff^\prime)LL}_{ijkl}\, \overline{f}_{Li} \gamma^\mu f_{L j}\,\overline{f^\prime}_{L k} \gamma_\mu f^\prime_{L l} + C^{(ff^\prime)RR}_{ijkl}\, \overline{f}_{R i} \gamma^\mu f_{R j}\,\overline{f^\prime}_{R k} \gamma_\mu f^\prime_{R l}\,\nonumber \\
 &+& C^{(ff^\prime)LR}_{ijkl}\, \overline{f}_{L i} \gamma^\mu f_{L j}\,\overline{f^\prime}_{R k} \gamma_\mu f^\prime_{R l} + C^{(ff^\prime)RL}_{ijkl}\, \overline{f}_{R i} \gamma^\mu f_{R j}\,\overline{f^\prime}_{L k} \gamma_\mu f^\prime_{L l}\,,\eeqa
where
\be \label{C}
C^{(ff^\prime)P P^\prime}_{ijkl} = \frac{g_F^2}{8 M_b^2}\,{\cal R}_{ab} {\cal R}_{cb}\, \left(\tilde{\lambda}^a_{f}{}_P \right)_{ij} \left(\tilde{\lambda}^c_{f^\prime}{}_{P^\prime} \right)_{kl}\,,\ee
and $P,P^\prime = L,R$ and $f, f^\prime = u, d, e$. The coefficients of the effective operators can be further simplified for the hierarchical gauge boson mass spectrum. Using Eqs. (\ref{MGB_block},\ref{R_block}), we find at leading order in $\rho$
\beqa \label{C_1}
\frac{8}{g_F^2} C^{(ff^\prime)P P^\prime}_{ijkl} &\simeq & \frac{1}{M_\alpha^2} \left( (R_3)_{\beta \alpha} (R_3)_{\gamma \alpha}  \left(\tilde{\lambda}^\beta_f{}_P \right)_{ij} \left(\tilde{\lambda}^\gamma_{f^\prime}{}_{P^\prime} \right)_{kl} \right. \nonumber \\
&+& \left.  (R_3)_{\beta \alpha} (\rho^T R_3)_{m \alpha}  \left(\tilde{\lambda}^\beta_{f}{}_P \right)_{ij} \left(\tilde{\lambda}^m_{f^\prime}{}_{P^\prime} \right)_{kl} +  (\rho^T R_3)_{m \alpha} (R_3)_{\beta \alpha}  \left(\tilde{\lambda}^m_{f}{}_P \right)_{ij} \left(\tilde{\lambda}^\beta_{f^\prime}{}_{P^\prime} \right)_{kl}  \right) \nonumber \\
& + & \frac{1}{M_m^2} \left( (R_5)_{nm} (R_5)_{pm}  \left(\tilde{\lambda}^n_{f}{}_P \right)_{ij} \left(\tilde{\lambda}^p_{f^\prime}{}_{P^\prime} \right)_{kl} \right. \nonumber \\
&-& \left.  (\rho R_5)_{\alpha m} (R_5)_{nm}  \left(\tilde{\lambda}^\alpha_{f}{}_P \right)_{ij} \left(\tilde{\lambda}^n_{f^\prime}{}_{P^\prime} \right)_{kl} -  (R_5)_{nm} (\rho R_5)_{\alpha m}  \left(\tilde{\lambda}^n_{f}{}_P \right)_{ij} \left(\tilde{\lambda}^\alpha_{f^\prime}{}_{P^\prime} \right)_{kl}  \right)\,.\eeqa

Further simplification can be achieved in the explicit model with the help of Eqs. (\ref{R_res}) and (\ref{MZ1-MZ2}). Substituting them in the above, we find
\beqa \label{C_2}
\frac{8 M_{Z_1}^2}{g_F^2}\, C^{(ff^\prime) P P^\prime}_{ijkl} &\simeq &  \sum_{\alpha=1}^3 \left(\tilde{\lambda}^\alpha_{f}{}_P \right)_{ij} \left(\tilde{\lambda}^\alpha_{f^\prime}{}_{P^\prime} \right)_{kl} + \epsilon^2 \sum_{m=4}^7 \left(\tilde{\lambda}^m_{f}{}_P \right)_{ij} \left(\tilde{\lambda}^m_{f^\prime}{}_{P^\prime} \right)_{kl} \nonumber \\
& - & \frac{\sqrt{3}}{4} \epsilon^2 \left( \left(\tilde{\lambda}^3_{f}{}_P \right)_{ij} \left(\tilde{\lambda}^8_{f^\prime}{}_{P^\prime} \right)_{kl}  + \left(\tilde{\lambda}^8_{f}{}_P \right)_{ij} \left(\tilde{\lambda}^3_{f^\prime}{}_{P^\prime} \right)_{kl} \right) \nonumber \\
& + & \frac{3}{4} \epsilon^2 \left(\tilde{\lambda}^8_{f}{}_P \right)_{ij} \left(\tilde{\lambda}^8_{f^\prime}{}_{P^\prime} \right)_{kl} + {\cal O} (\epsilon^4)\,.\eeqa
At the leading order, the flavour violation is governed by the coupling matrices $\tilde{\lambda}^\alpha_f{}_{L,R}$. Since they do not commute with each other, all of them cannot take diagonal form for any $U_{L,R}$. Therefore, the most dominant flavour violations in the model are captured by the coefficients
\beqa \label{C_3}
C^{(ff^\prime) P P^\prime}_{ijkl} &\simeq & \frac{g_F^2}{8 M_{Z_1}^2}  \sum_{\alpha=1}^3 \left(\tilde{\lambda}^\alpha_{f}{}_P \right)_{ij} \left(\tilde{\lambda}^\alpha_{f^\prime}{}_{P^\prime} \right)_{kl}\,.\eeqa
The above can be used to estimate the most dominant contribution to various flavour-violating processes in the quark and lepton sectors.

\subsection{Quark sector}
The strongest constraints on various $C^{(ff^\prime) P P^\prime}_{ijkl}$ primarily arise from meson-antimeson oscillations such as $K^0-\overline{K}^0$, $B_d-\overline{B}_d^0$, $B_s-\overline{B}_s^0$ and $D^0-\overline{D}^0$. To quantify these constraints, we closely follow the procedure adopted in our previous analysis \cite{Mohanta:2022seo}.  Comparing Eq. (\ref{d6_ops}) with the effective Hamiltonian, ${\cal H}_{eff}\,=\, \sum_{i=1}^5 C_M^i Q^i + \sum_{i=1}^3 \tilde{C}_M^i \tilde{Q}_i$, which parametrizes $\Delta F = 2$ transitions $M-\overline{M}$ \cite{UTfit:2007eik}, we find the effective Wilson coefficients at $\mu=M_{Z_1}$ as
\begin{align}
C^1_K \,=\, -C^{(dd)LL}_{1212} &,& \tilde{C}^1_K= -C^{(dd)RR}_{1212} &,&{C}^5_K= -4\, C^{(dd)LR}_{1212}\,,\\
C^1_{B_d} \,=\, -C^{(dd)LL}_{1313} &,& \tilde{C}^1_{B_d}= -C^{(dd)RR}_{1313} &,&{C}^5_{B_d}= -4\, C^{(dd)LR}_{1313}\,,\\
C^1_{B_s} \,=\, -C^{(dd)LL}_{2323} &,& \tilde{C}^1_{B_s}= -C^{(dd)RR}_{2323} &,&{C}^5_{B_s}= -4\, C^{(dd)LR}_{2323}\,,\\
C^1_{D} \,=\, -C^{(uu)LL}_{1212} &,& \tilde{C}^1_D= -C^{(uu)RR}_{1212} &,& {C}^5_D= -4\, C^{(uu)LR}_{1212}\,.
\end{align} 
The remaining $C^i_M$ and $\tilde{C}^i_M$ are vanishing at this scale.

Using the renormalization group equations, we evolve all the coefficients from $\mu=M_{Z_1}$ to the $\mu = 2$ GeV for $K$ meson \cite{Ciuchini:1998ix}, $\mu = 4.6$ GeV for $B$ mesons  \cite{Becirevic:2001jj} and $\mu = 2.8$ GeV for $D$ meson \cite{UTfit:2007eik}. The running gives rise to non-vanishing $C^4_M$ while $C^{2,3}_M$ and $\tilde{C}^{2,3}_M$ remains zero. The evolved Wilson coefficients are computed using Eq. (\ref{C_3}) for the three benchmark solutions listed in Table \ref{tab:sol} and are compared with the corresponding experimental limits obtained by the UTFit collaboration \cite{UTfit:2007eik}. The results are listed in Table \ref{tab:meson_WC}. It can be seen that the strongest limits arise from $K^0-\overline{K}^0$ mixing which disfavours $M_{Z_1} \leq 10^6$ GeV. The same limit was also observed in our previous framework based on flavour non-universal abelian symmetries. 
\begin{table}[t]
\begin{center}
\begin{tabular}{ccccc} 
\hline
\hline
~~Wilson coefficient ~~&~~Allowed range~~&~~S1~~&~~S2~~&~~S3~~\\
 \hline
Re$C_K^1$ & $[-9.6,9.6]\times 10^{-13}$ & \rb{$ 6.0\times 10^{-11}$} & \gb{$-1.0 \times 10^{-14} $} & \gb{$ -5.2 \times 10^{-19}$}\\
Re$\tilde{C}_K^1$ & $[-9.6,9.6]\times 10^{-13}$ & \gb{$ -4.8\times 10^{-16}$}& \gb{$ -1.7\times 10^{-19}$} & \gb{$4.0 \times 10^{-24}$}\\
Re$C_K^4$ & $[-3.6,3.6]\times 10^{-15}$ & \rb{$ 3.2 \times 10^{-10}$} & \rb{$-1.0 \times 10^{-13}$} & \gb{$ -2.5 \times 10^{-18}$} \\
Re$C_K^5$ & $[-1.0,1.0]\times 10^{-14}$ & \rb{$ 2.8 \times 10^{-10}$} & \rb{$ -8.3 \times 10^{-14}$} & \gb{$ -2.0 \times 10^{-18}$}\\
Im$C_K^1$ & $[-9.6,9.6]\times 10^{-13}$ & \rb{$ 4.4 \times 10^{-11}$}& \gb{$ 4.3\times 10^{-15}$} & \gb{$ -3.8 \times 10^{-19}$}\\
Im$\tilde{C}_K^1$ & $[-9.6,9.6]\times 10^{-13}$ &\gb{$ -3.4 \times 10^{-15}$} & \gb{$-5.8 \times 10^{-19}$} & \gb{$1.8  \times 10^{-23}$}\\
Im$C_K^4$ &  $[-1.8,0.9]\times 10^{-17}$  & \rb{$ -1.6 \times 10^{-10}$}  & \rb{$ -4.3 \times 10^{-14}$} &\gb{$ 1.4 \times 10^{-18}$} \\
Im$C_K^5$ & $[-1.0,1.0]\times 10^{-14}$ &\rb{$  -1.4\times 10^{-10}$} & \rb{$ -3.6 \times 10^{-14}$} & \gb{$ 1.1 \times 10^{-18}$}\\
\hline
$|C_{B_d}^1|$ & $<2.3\times 10^{-11}$ & \rb{$ 9.2\times 10^{-11}$} & \gb{$ 1.2\times 10^{-14}$} & \gb{$ 8.0 \times 10^{-19}$} \\
$|\tilde{C}_{B_d}^1|$ & $<2.3\times 10^{-11}$  & \gb{$ 2.2\times 10^{-12}$} & \gb{$ 2.8\times 10^{-16}$} & \gb{$ 1.6\times 10^{-20}$}\\
$|C_{B_d}^4|$ &  $<2.1\times 10^{-13}$ &  \rb{$ 9.2 \times 10^{-11}$}   & \gb{$ 1.4\times 10^{-14}$} & \gb{$8.1 \times 10^{-19}$}\\
$|C_{B_d}^5|$ & $<6.0\times 10^{-13}$  & \rb{$ 1.6 \times 10^{-10}$} & \gb{$ 2.4\times 10^{-14}$} & \gb{$ 1.3\times 10^{-18}$}\\
\hline
$|C_{B_s}^1|$ & $< 1.1 \times 10^{-9}$ & \gb{$ 1.1  \times 10^{-11 }$} & \gb{$ 2.9\times 10^{-15 }$} & \gb{$ 7.7\times 10^{ -20}$}\\
$|\tilde{C}_{B_s}^1|$ & $< 1.1 \times 10^{-9}$ & \gb{$ 1.7\times 10^{-12}$} & \gb{$ 2.3\times 10^{-16}$} & \gb{$1.3 \times 10^{-20}$}\\
$|C_{B_s}^4|$ & $< 1.6 \times 10^{-11}$ & \gb{$ 1.3 \times 10^{-11}$}  & \gb{$ 2.9\times 10^{-15}$} & \gb{$ 1.1\times 10^{-19}$} \\
$|C_{B_s}^5|$ & $< 4.5 \times 10^{-11}$ & \gb{$ 2.4 \times 10^{-11}$} & \gb{$ 4.8\times 10^{-15}$} & \gb{$  1.8\times 10^{-19}$}\\
\hline
$|C_D^1|$ & $<7.2 \times 10^{-13}$ & \rb{$ 5.5\times 10^{-11 }$} & \gb{$ 7.5\times 10^{-15 }$} & \gb{$6.5 \times 10^{ -19}$}\\
$|\tilde{C}_D^1|$ & $<7.2 \times 10^{-13}$ & \gb{$ 4.4 \times 10^{ -16}$} & \gb{$ 5.6\times 10^{-20}$} & \gb{$ 5.4\times 10^{-24}$}\\
$|C_D^4|$ & $<4.8\times 10^{-14}$ & \rb{$ 1.9\times 10^{-10}$} & \gb{$ 3.6\times 10^{-14}$} & \gb{$ 2.1\times 10^{-18}$} \\
$|C_D^5|$ & $<4.8 \times 10^{-13}$ & \rb{$ 2.2\times 10^{-10}$} & \gb{$ 4.1\times 10^{-14}$} & \gb{$2.2 \times 10^{-18}$}\\
\hline
\hline
\end{tabular}
\end{center}
\caption{Numerical values of various Wilson coefficients (in ${\rm GeV}^{-2}$ unit) of the operators leading to meson-antimeson oscillations estimated for three example solutions. The experimentally allowed ranges are taken from \cite{UTfit:2007eik}. The values highlighted in red are excluded by the respective limits.}
\label{tab:meson_WC}
\end{table}

\subsection{Lepton sector}

As noted in the previous section, the dominant contribution to the flavour violation process is governed by the first three gauge bosons $B^\alpha_\mu $. The exchange of $B^\alpha_\mu $ mediate lepton flavour violating process like $\mu \to e$ conversion in nuclei, $l_i \to 3 l_j $ and $l_i \to \l_j \gamma$. The first two processes arise at the tree level whereas the latter is at the one-loop level in the present model. 

The $\mu \to e$ conversion in the field of the nucleus is strongly constrained by the SINDRUM II experiment \cite{SINDRUMII:2006dvw} which uses ${}^{197}Au$ nucleus. The relevant branching ratio estimated in \cite{Kitano:2002mt} is given by
\be \label{mu2e}
{\rm BR}[\mu \to e] = \frac{2 G_F^2}{\omega_{\rm capt}}\,(V^{(p)})^2\, \left(|g^{(p)}_{LV}|^2 + |g^{(p)}_{RV}|^2\right)\,. \ee
Here, $V^{(p)} =  0.0974\, m_\mu^{5/2}$ is an integral  involving proton distribution and ${\omega_{\rm capt}}  = 13.07 \times 10^{6}\,{\rm s}^{-1}$ is muon capture rate for ${}^{197}Au$  \cite{Kitano:2002mt}.  $g^{(p)}_{LV, RV}$  depend on the flavour-violating couplings and they are parametrized as
\be \label{gLV}
g^{(p)}_{LV,RV} = 2 g^{(u)}_{LV,RV} + g^{(d)}_{LV,RV}\,.\ee
The expressions for $g^{(q)}_{LV,RV}$ ($q=u,d$) can be obtained using Eq. (\ref{d6_ops}) and  \cite{Smolkovic:2019jow}. For the present model,  we find
\beqa \label{gLV_Z1}
g^{(q)}_{LV} &\sim & \frac{\sqrt{2}}{G_F} \, \frac{1}{2}
\left[C^{(eq) L L}_{2111}\,+\,C^{(eq) L R}_{2111} \right]\,\nonumber \\
g^{(q)}_{RV} &\sim & \frac{\sqrt{2}}{G_F} \, \frac{1}{2}
\left[C^{(eq) R R}_{2111}\,-\,C^{(eq) R L}_{2111} \right]\, .\eeqa
The branching ratios for $\mu$ to $e$ conversion, computed using Eqs. (\ref{mu2e}, \ref{gLV}) and (\ref{gLV_Z1}), for the three benchmark solutions are given in Table \ref{tab:lfv}.  The present experimental  limit disfavours S1  in this case.
\begin{table}[t]
\begin{center}
\begin{tabular}{ccccc}
\hline
\hline
~~LFV observable~~&~~Limit~~&~~S1~~&~~S2~~&~~S3~~\\
\hline
${\rm BR}[\mu \to e]$ & $< 7.0 \times 10^{-13}$ & \rb{$ 1.0 \times 10^{-8}$} & \gb{$ 1.4 \times 10^{-16 }$} & \gb{$ 1.4 \times 10^{-24 }$}\\
\hline
${\rm BR}[\mu \to 3e]$ & $< 1.0 \times 10^{-12}$  & \rb{$2.4 \times 10^{-11 }$} & \gb{$5.0  \times 10^{-19 }$} & \gb{$ 1.6 \times 10^{-27 }$}\\
${\rm BR}[\tau \to 3\mu]$ & $< 2.1 \times 10^{-8}$ & \gb{$2.3 \times 10^{-11 }$} & \gb{$4.2 \times 10^{-19 }$}& \gb{$ 1.7 \times 10^{-27 }$}\\
${\rm BR}[\tau \to 3 e]$ & $< 2.7 \times 10^{-8}$ & \gb{$9.4 \times 10^{-12}$} & \gb{$ 2.7\times 10^{-19 }$}& \gb{$ 6.0 \times 10^{-28 }$}\\
\hline
${\rm BR}[\mu \to e \gamma]$ & $< 4.2 \times 10^{-13}$ & \rb{$ 7.0\times 10^{-9 }$} & \gb{$ 4.9\times 10^{-17 }$}& \gb{$ 6.8 \times 10^{-25 }$}\\
${\rm BR}[\tau \to \mu \gamma]$ &  $< 4.4 \times 10^{-8}$ & \gb{$2.1 \times 10^{-11 }$}& \gb{$1.6 \times 10^{-19 }$}& \gb{$2.2  \times 10^{-27 }$}\\
${\rm BR}[\tau \to e \gamma]$ &  $< 3.3 \times 10^{-8}$ & \gb{$ 1.3\times 10^{ -12}$} & \gb{$ 9.3\times 10^{-21 }$}& \gb{$1.3  \times 10^{-28}$}\\
\hline
\hline
\end{tabular}
\end{center}
\caption{Branching ratios evaluated for various charged lepton flavour violating processes for the three benchmark solutions  listed in Table \ref{tab:sol}. The corresponding  experimental limits are extracted from \cite{Calibbi:2017uvl}.  The values excluded by the limits are highlighted in red.}
\label{tab:lfv}
\end{table}

The trilepton decay, $l_i \to 3 l_j $, is mediated by the new gauge bosons at the tree-level. The decay width for this process can be estimated following \cite{Ramond:1999vh}. For the present model and in the limit $m_i \gg m_j$, we find
\beqa \Gamma[l_i \to 3 l_j] &=&  \frac{4 \, m_i^5}{1536}\, \sum_{P,P'}\left|
C^{(ee) P P^\prime}_{jijj}\right|^2\,,\eeqa
which, using Eq. (\ref{C_3}), takes the following form
\beqa
\label{l_i3lj}\Gamma[l_i \to 3 l_j] \,=\, \frac{g_F^4}{16}\frac{ m_i^5}{1536}\, \sum_{\alpha,\beta=1}^3 &{}& \left[ \left(\tilde{\lambda}^\alpha_{e}{}_L \right)_{ji} \left(\tilde{\lambda}^\beta_{e}{}_L \right)_{ji} +  \left(\tilde{\lambda}^\alpha_{e}{}_R \right)_{ji} \left(\tilde{\lambda}^\beta_{e}{}_R \right)_{ji} \right] \, \nonumber\\
& & \times \left[ \left(\tilde{\lambda}^\alpha_{e}{}_L \right)_{jj} \left(\tilde{\lambda}^\beta_{e}{}_L \right)_{jj} +  \left(\tilde{\lambda}^\alpha_{e}{}_R \right)_{jj} \left(\tilde{\lambda}^\beta_{e}{}_R \right)_{jj} \right]\, . 
\eeqa
The branching ratios for  $\mu \to 3 e $, $\tau \to 3 e $ and $\tau \to 3 \mu $  evaluated using the above expression are given in Table \ref{tab:lfv} for three solutions along with their corresponding experimental limits.

To estimate the branching ratios for $l_i \to \l_j \gamma$, we follow \cite{Lavoura:2003xp} and compute the decay width in the approximation $M_{Z_1} \gg m_i, m_j$. The result can be parametrized as
\be \label{i2j_gamma} \Gamma[l_i \to \l_j \gamma] \,\simeq\, \frac{\alpha {g_F}^4}{64} \left(1-\frac{m_j^2}{m_i^2}\right)^3\, {m_i^5}\,\left( \left|\sigma_L \right|^2\,+\, \left|\sigma_R \right|^2\right)\,,\ee
where
\be \sigma_L \,=\, \sum_{\alpha=1}^{3} \sum_{k=1}^{3} \left[Y_1\, \frac{m_j}{m_i}\left(\tilde{\lambda}^\alpha_L \right)_{jk}\left(\tilde{\lambda}^\alpha_L \right)_{ki} \, +\,Y_2\,\left(\tilde{\lambda}^\alpha_R \right)_{jk}\left(\tilde{\lambda}^\alpha_R \right)_{ki}\,
- 4 Y_3\,\frac{m_k}{m_i}\,\left(\tilde{\lambda}^\alpha_R \right)_{jk}\left(\tilde{\lambda}^\alpha_L \right)_{ki}\right]\, . \ee
Similarly, $\sigma_R$ can be obtained by replacing $L \leftrightarrow  R$ in the above expression.  $Y_1, Y_2$ and $Y_3$ are loop functions and they are given by
\be Y_1 = Y_2 = 2a+6c+3d\, ~~\,~~
Y_3 = a + 2c\,, \ee
and the explicit expressions of $a, c$, and $d$ are given in \cite{Lavoura:2003xp}. Using Eq. (\ref{i2j_gamma}), we estimate the branching ratios for $\mu \to  e  \gamma$, $\tau \to \mu \gamma$ and $\tau  \to e \gamma $  for three benchmark solutions and list them in Table \ref{tab:lfv}.

Comparing the estimated magnitudes of various charged lepton flavour violating observables given in Table \ref{tab:lfv} with the corresponding experimental limits, we find that $M_{Z_1}\leq 10^{4}$ GeV are excluded. However, this seems to be a much weaker constraint compared to the  one arising from the quark flavour-violating process. Altogether the strongest limit on the scale of $SU(3)_F$ breaking comes from $K$-$\overline{K}$  mixing which disfavours $M_{Z_1} \le 10^3$ TeV. 

The flavour constraints on the new physics in this class of models are dominant and they supersede the other limits put by direct searches or precision electroweak observables as shown by us in our previous work \cite{Mohanta:2022seo}. For example, the strongest limit from the direct searches at the LHC  implies $M_{Z_1} > 7.20$ TeV \cite{CMS:2018mgb}. Similar constraints  on the vectorlike quarks,  $m_B > 1.57$ TeV \cite{ATLAS:2018mpo,CMS:2020ttz} and $m_T > 1.31$ TeV \cite{CMS:2018wpl,ATLAS:2018ziw}, are even more weaker. The $Z_{1,2}$ bosons can mix with the SM $Z$ boson through VEVs of  $H_{ui}$ and $H_{di}$ in the present model which in turn contributes to the electroweak observables.  The most  stringent limits in this case also imply $M_{Z_1} \geq  4.5$  TeV \cite{Mohanta:2022seo} making all these constraints irrelevant in comparison to the ones originating from the quark and lepton flavour violations. 

\section{Conclusion}
\label{sec:concl}
We have discussed a mechanism for generating loop-induced masses for the first and second generations of quarks and charged leptons using a gauged horizontal $SU(3)_F$ symmetry. The field content of the theory ensures that the Yukawa sector has an accidental global symmetry leading to vanishing masses for lighter generations of fermions. This symmetry is broken by the $SU(3)_F$ gauge interactions which then radiatively induces the masses for the otherwise massless fermions. We find that the radiative corrections typically generate masses for both the second and first-generation fermions at 1-loop. The hierarchy between the two, therefore, requires a separate explanation. We show that this is possible if $SU(3)_F$ is broken in two steps with an intermediate $SU(2)$ symmetry. This leads to a little hierarchy among the gauge bosons of the local flavour group which is then transferred to the fermion sector through quantum corrections. We construct an explicit model based on this mechanism and show how the hierarchical quark and lepton masses can be viably reproduced.

A similar setup based on flavour non-universal $U(1) \times U(1)$ symmetry was proposed by us earlier in \cite{Mohanta:2022seo}. The framework presented in this paper replaces the pair of non-abelian symmetries with $SU(3)_F$ leading to two important improvements. The gauge boson mass hierarchy, which was an ad-hoc assumption in the case of $U(1) \times U(1)$, now emerges from a sequential breaking of single gauge group $SU(3)_F$. Secondly, the non-abelian single flavour group leads to a more predictive framework in terms of the number of Yukawa couplings in the theory. The number of free yukawa couplings after removing the unphysical phases reduces from 20 in \cite{Mohanta:2022seo} to 12 in the present model. This implies correlations among the masses of various quarks and charged leptons. An example of this is seen in the specific model which favours the value of strange quark mass $2.4 \sigma$ smaller than the present experimental value.

Phenomenologically, the $SU(3)_F$ breaking scale is constrained from below entirely from the flavour violation. The new gauge bosons have ${\cal O}(1)$ flavour-changing couplings with fermions leading to large rates for flavour-violating processes. This feature seems to be inherently present in the frameworks of radiative mass generation mechanisms. We estimate various quark and lepton flavour-violating observables and find that the lightest gauge boson of $SU(3)_F$ is required to be heavier than $10^3$ TeV as implied by the present limits. This makes it impossible to verify such a framework in the direct search experiments. Nevertheless, the specific model can still be probed indirectly through precision measurements of fermion masses and mixing parameters and flavour-violating observables.

\section*{Acknowledgements}
This work is partially supported under the MATRICS project (MTR/2021/000049) by the Science \& Engineering Research Board (SERB), Department of Science and Technology (DST), Government of India. KMP acknowledges support from the ICTP through the Associates Programme (2023-2028) where part of this work was completed and preliminary results were presented.

\appendix

\section{$SU(3)_F$ generators}
\label{app:GM}
An explicit form of the $SU(3)_F$ generators $\lambda^a$ (with $a=1,...,8$) that we use in the present work is 
\begin{align*}
\lambda^1 &= \left( \ba{ccc} 0 & 0 & 0 \\ 0 & 0 & 1\\0&1&0 \ea\right)\,,\, \lambda^2 \,=\, \left( \ba{ccc} 0 & 0 & 0 \\ 0 & 0 & -i\\0&i&0 \ea\right)\,,\, & \lambda^3 \,=\, \left(\ba{ccc} 0 & 0& 0\\ 0& 1 &0\\0 & 0 & -1 \ea \right)\,,\, & ~~\lambda^4 \,=\, \left(\ba{ccc} 0 & 1& 0\\ 1& 0 &0\\0 & 0 & 0 \ea \right)\,,\\
 \lambda^5 \,&=\, \left(\ba{ccc} 0 & i& 0\\ -i& 0 &0\\0 & 0 & 0 \ea \right)\,,\, 
\lambda^6 \,=\, \left(\ba{ccc} 0 & 0& 1\\ 0& 0 & 0\\1 & 0 & 0 \ea \right)\,,\, &\lambda^7 \,=\, \left(\ba{ccc} 0 & 0& i\\ 0& 0 &0\\-i & 0 & 0 \ea \right)\,,\, & ~\lambda^8 \,=\,\frac{1}{\sqrt{3}} \left(\ba{ccc} -2 & 0& 0\\ 0& 1 &0\\0 & 0 & 1 \ea \right)\,.\end{align*}

These  are written in the basis such that the first three generators correspond to the gauge bosons which do not couple to the first generation in the canonical basis. An $SU(2)$  subgroup corresponding to these three generators of the full flavour symmetry group  remains unbroken by the VEV of $\eta_1$ in Eq. (\ref{eta_vevs}).

\section{Scalar Potential and VEVs}
\label{app:potential}
In this Appendix, we demostrate the conditions which lead to the VEV configurations of $\eta_{1,2}$ fields considered in Eq. (\ref{eta_vevs}). Since the flavour symmetry breaking scale is required to be much larger than the electroweak scale, we neglect the small contribution that may arise from the VEVs of $H_{u,d}^i$ in the $SU(3)_F$ breaking. The most general and renormalizable potential involving $\eta_{1,2}$ can be written as
\beqa \label{pot}
V(\eta_1,\eta_2) &=& m_{11}^2\, \eta_1^\dagger \eta_1 + m_{22}^2\, \eta_2^\dagger \eta_2 - \left\{ m_{12}^2\, \eta_1^\dagger \eta_2 + {\rm h.c.} \right\} \nonumber \\
 &+& \frac{\xi_1}{2}\, (\eta_1^\dagger \eta_1)^2 +  \frac{\xi_2}{2}\, (\eta_2^\dagger \eta_2)^2 + \xi_3\,  (\eta_1^\dagger \eta_1) (\eta_2^\dagger \eta_2) + \xi_4\,  (\eta_1^\dagger \eta_2) (\eta_2^\dagger \eta_1) \nonumber \\
 &+& \left\{\frac{\xi_5}{2} (\eta_1^\dagger \eta_2)^2 + \xi_6\,  (\eta_1^\dagger \eta_1) (\eta_1^\dagger \eta_2) + \xi_7\, (\eta_2^\dagger \eta_2) (\eta_1^\dagger \eta_2) + {\rm h.c.} \right\}\,.\eeqa
 Here, all the parameters except $\xi_{5,6,7}$ and $m_{12}^2$ are real. 
 
For the VEV configuration of $\eta_{1,2}$ given in Eq. (\ref{eta_vevs}), the minimization of the potential implies 
\beqa \label{minimum}
v\, (m_{11}^2 + v^2 \xi_1 + \epsilon^2 v^2 \xi_3) &=& 0 \,, \nonumber \\
\epsilon v\, (m_{22}^2 + \epsilon^2 v^2 \xi_2 + v^2 \xi_3) &=& 0\,.\eeqa
The non-trivial solution of the above equations corresponds to
\be \label{vevs_expl}
v^2 = \frac{-m_{11}^2 \xi_2 + m_{22}^2 \xi_3}{\xi_1 \xi_2 - \xi_3^2}\,,~~~(\epsilon v)^2 = \frac{-m_{22}^2 \xi_1 + m_{11}^2 \xi_3}{\xi_1 \xi_2 - \xi_3^2}\,.\ee
The VEVs are obtained in terms of real parameters $m_{11}^2$, $m_{22}^2$ and $\xi_{1,2,3}$. The latter are also constrained by the stability of potential:
\be \label{stability}
\xi_{1,2} \geq 0\,,~~\xi_3 \geq - \sqrt{\xi_1 \xi_2}\,. \ee

For $0 > \xi_3 \ge -\sqrt{\xi_1 \xi_2}$, one finds $ \xi_1 \xi_2 - \xi_3^2 \ge 0$. Further assuming $|m_{22}^2| \ll |m_{11}^2|$, $\xi_1 \ll \xi_2$ and $m_{11}^2 < 0$, it can be seen that the VEVs in Eq. (\ref{vevs_expl}) are real. Their ratio is then determined as
\be \label{ratapp}
\epsilon^2 \approx -\frac{\xi_3}{\xi_2} \le \sqrt{\frac{\xi_1}{\xi_2}} \ll 1\,.\ee
Moreover, for $\xi_3 \approx -\sqrt{\xi_1 \xi_2}$, the VEVs obtained Eq. (\ref{vevs_expl}) turn out to be the global minima of the potential among the available solutions offered by Eq. (\ref{minimum}). In summary, one can obtain the desired VEV configurations for $\eta_{1,2}$ consistent with stability constraints for a specific choice of parameters.

As mentioned in the beginning, we have neglected contribution to the $SU(3)_F$ breaking from the scalars  charged also under the electroweak symmetry since the viable generation of fermion mass hierarchy along with the flavour constraints require at least four orders of magnitude separation between the two scales. This hierarchy among the scales, however, requires a fine-tuning. Since the terms like $\eta_a^\dagger \eta_a H_{u,d}^\dagger H_{u,d}$ would induce large bare mass terms for $H_{u,d}$ when $SU(3)_F$ gets broken, the VEVs of $H_{u,d}$ would naturally tend to stay close to the flavour symmetry breaking scale. Also, as the terms like $\eta_a^\dagger \eta_a H_{u,d}^\dagger H_{u,d}$ cannot be forbidden by any gauge or global symmetries within this non-supersymmetric framework, the seperation between the two scales is not stable under the quantum corrections and technically unnatural. This is similar to the usual gauge hierarchy problem and requires fine-tuning.

\bibliography{references}

\end{document}